# Dynamic Field Programmable Logic-Driven Soft Exosuit

**Frances Cleary**[1,2*], **Witawas Srisa-an**[4], **David C. Henshall**[1,3], **Sasitharan Balasubramaniam** [4]

[1]Physiology & Medical Physics, Royal College of Surgeons Ireland, Dublin, Ireland
[2]Walton Institute, South East Technological University, Waterford, Ireland
[3]FutureNeuro, The SFI Research Centre for Chronic and Rare Neurological Diseases, RCSI University of Medicine and Health Sciences, Dublin, Ireland
[4]School of Computing, University of Nebraska, Lincoln, United States

**Correspondence**: Corresponding Author francescleary20@rcsi.ie

**Abstract**: The next generation of etextiles foresees an era of smart wearable garments where embedded seamless intelligence provides the ability to sense, process and perform. Core to this vision is embedded textile functionality enabling dynamic configuration. In this paper we detail a methodology, design and implementation of a dynamic field programmable logic-driven fabric soft exosuit. Dynamic field programmability allows the soft exosuit to alter its functionality and adapt to specific exercise programs depending on the wearers need. The dynamic field programmability is enabled through motion based control arm movements of the soft exosuit triggering momentary sensors embedded in the fabric exosuit at specific joint placement points (right arm: wrist, elbow).The embedded circuitry in the fabric exosuit is implemented using a layered and interchangeable approach. This includes logic gate patches (AND,OR,NOT) and a layered textile interconnection panel. This modular and interchangeable design enhances the soft exosuits flexibility and adaptability. A truth table aligning to a rehabilitation healthcare use case was utilised. Tests were completed validating the field programmability of the soft exosuit and its capability to switch between its embedded logic driven circuitry and its operational and functionality options controlled by motion movement of the wearers right arm (elbow and wrist). Iterative exercise movement and acceleration based tests were completed to validate the functionality of the field programmable logic driven fabric exosuit. We demonstrate a working soft exosuit prototype with motion controlled operational functionality that can be applied to rehabilitation applications.

## Introduction

Smart garments bring the capability to collect, store and process data from an environment or from then body providing applications and products across multiple sectors [1] [2]. A fusion of textiles and technologies are being researched and investigated to progress new innovations and meet current challenges and limitations in this space [3] [4]. There is an increased interest in smart wearable garment applications providing remote healthcare monitoring, enabling early intervention before symptoms escalate [5] [6] [7] [8] [9]. Field programmability (FP) provides end users with reconfigurable hardware functionality meeting specific use case needs [10] [11].

Driven by the need for low cost prototypes, with low risk and high return, FP technologies enable configurable logic and interconnections allowing interchangeable operation. Currently, FP driven technology has been adopted and implemented across varying sectors, for example through ongoing research around the use of FP biological circuits for the computational design of configurable bio(logic) blocks [12] or research investigating hardware technologies such as field programmable gate array (FPGA) in order to develop deep network accelerators suitable for biomedical and healthcare applications [13]. In parallel researchers have been experimenting with the programmable nature of textiles, how microelectronics, artificial intelligence and new emerging technologies can be embedded in a textile environment to produce smarter, re-configurable and more impactful applications [14] [15]

[16] [17] [18]. Promoting the ambition of computing fabrics Loke et al convey fibre-based innovative possibilities through fabric computing and artificial intelligence (AI) [19].

Such a vision through real time wearable body surface signal data AI processing (acoustic, optical, biological) bring the promise of a disruptive era with the emerging next generation of fabric-AI driven etextiles. Further research in the etextiles domain through the application of dynamically re-configurable, interconnecting and intelligent embedded textile techniques is required to stem new textile fabric intelligent applications. A growth in the etextiles market size is expected to increase from 2.98 billion Dollars (2022) to 8.59 billion Dollars (2028) [20].

In this paper we design and develop a dynamic field In this paper we design and develop a dynamic field programmable logic-driven fabric soft exosuit. The approach takes inspiration from combinational logic circuits, field programmable digital techniques and embedded textile intelligence. We implement and validate a smart fabric soft exosuit with the capability to alter its embedded combinational logic circuit configuration, hence modifying the soft exosuits embedded intelligence and its end user application use cases. The soft exosuits layered inter-operable circuitry adopts a flexible design with a modular and interchangeable fabric panel patch based approach enabling re-usability through easy application modification, hence, also providing enhanced garment sustainability. Validation of the dynamic soft exosuit included testing the functionality of the FP combinational logic circuit, testing the individual logic gates operational status as well as the Dynamic FP interchangeable embedded circuit switching capability. From test results obtained the operational logic of each circuit functioned as expected, as per the defined truth table when switching from circuit 1 to circuit 2, through motion triggered control movements of the elbow or wrist in the right arm of the soft exosuit. To showcase the usability of the dynamic FP soft exosuit, alignment to a selected post-stroke upper limb rehabilitation exercise programme was completed with a mapping to the Fugl-Meyer upper limb manual assessment scale method. Specific exercise based movements defined for the embedded and interchangeable circuit 1 and 2 are demonstrated successfully through an angle movement analysis. Application of the dynamic FP soft exosuit to task driven exercise scenarios is also presented, showing the extension of the soft exosuit to real world task driven exercises (lift bottle, hold towel, lift bag etc).

Dynamic field programmable textile garments have the potential to impact multiple advancing markets. We focus on the application and benefits of such a dynamic smart garment for the healthcare sector, where embedded dynamically changeable intelligence in a wearable garment environment has the ability to provide patients with a wearable means to aid required upper limb exercises.

The rest of the paper is organized as follows: Section 2 discusses the Materials and Methods adopted. Section 3 gives detail around the design and hardware implementation of the dynamic FP soft fabric exosuit. Section 4 gives a summary of the experimental results completed and section 5 details the application area example use cases with specific focus on healthcare post-stroke rehabilitation. Section 6 is the final conclusion section that concludes the paper and provides insights into future potential work.

## Materials and Methods

This section describes the research design methods and materials selected and utilised in order to implement the dynamic FP fabric soft exosuit. Key design methodology aspects considered and adopted include:

- A layered fabric based circuit design process and method for the creation of the dynamic FP embedded combinational logic circuit in the soft exosuit.

- Identification of a suitable modular method and process for the creation and interchangeability of logic gate patches and conductive thread based wired interconnections.
- Momentary textile sensors as rehabilitation exercise triggered actuators, also acting as enablers for dynamic field programmability in a fabric environment.
- Identification and definition of a healthcare stroke rehabilitation use case and workflow methodology suitable for the soft exosuit.

## Dynamic FP Fabric Soft Exosuit Circuit Functionality Methodology

A programmable combinational logic circuit design was constructed with the functionality to alternate dynamically between two combinational circuits in the one overall embedded fabric circuit in the soft exosuit as illustrated in (Fig 1(a-b)).

Each circuit once activated had different information processing operational functionality based on the initial assigned truth tables and logic gates linked to the chosen healthcare rehabilitation use case shown in (Fig 1(d)). The circuit dynamic field programmability and behaviour is triggered through control momentary textile sensors embedded in the wrist and elbow of the circuit controlling arm (right arm) of the soft exosuit which can be viewed in (Fig 1(c)). Key control motion movements of the soft exosuit include the following

- On the control arm (right arm) of the soft exosuit, when its (wrist) is in a flexion bending motion placement state, this activates the combinational logic (circuit 1) in the dynamic FP fabric exosuit. Cluster connections (2,4 ) from the momentary sensors in the control arm (wrist) to the combinational circuit embedded in the back panel of the soft exosuit are activated and the (circuit 1) logic truth table and defined exercises (Wrist Flexion and Elbow Flexion) are enabled for use by the wearer as illustrated in (Fig 1(a)) and (Fig 2(d)).
- On the control arm (right arm) of the soft exosuit, when its (elbow) is in a flexion bending motion placement state, this activates the combinational logic (circuit 2) in the dynamic FP fabric exosuit. Cluster connections (1,3,5) from the momentary sensors to the combinational circuit embedded in the back panel of the soft exosuit are activated and the (circuit 2) logic truth table and defined exercises (Thumb Flexion and Finger Flexion) are enabled for use by the wearer as seen in (Fig 1(a)) and (Fig 2(d)).

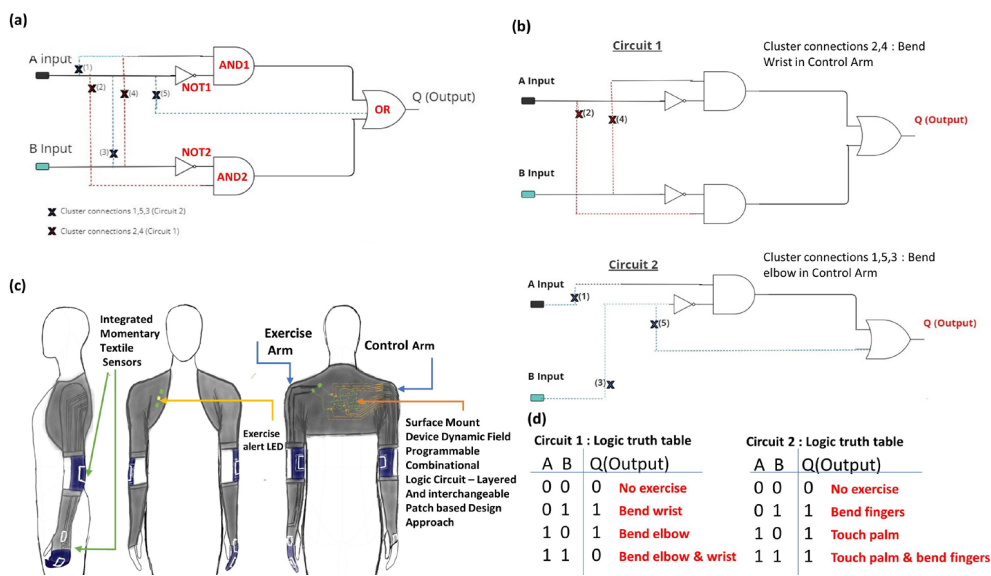

Fig. 1. a) Full combinational logic circuit with highlighted wired connection cluster break points activating circuit one or two, b) The two dynamic field programmable circuits capable of being switched within the textile garment environment,c)

Concept design of the smart shrug garment created during the methodology and materials analysis research phase d) Two defined logic truth tables that map to the separate circuits linking to the assigned use case exercises.

The dynamic alternation between the 2 embedded combinational circuits demonstrating the FP of the fabric exosuit can be viewed in Fig (2(d-e)). An initial calibration state of the soft exosuit to ensure no inputs are active and no circuit selection is enabled, involves both hands of the fabric exosuit being placed in a downwards straight position. To ensure that the output from the soft exosuit was in a zero (no input/ no output) state, a standard light emitting diode (LED) to visually highlight the output is utilised, where LED is ON equals a '1' state and where LED is OFF equals a '0' state. Hence this enables us to visually align the working functionality of the soft exosuit to the defined truth tables. The defined truth tables align to the selected exercise program and movements available to the wearer.

Fig 2(d) details the circuit activation based on the enabled control arm momentary sensors, in this case bending of the (wrist) to enable (circuit 1) (colour 'yellow' active (conductive thread)) interconnections in the image). Here we demonstrate the set inputs (0,1)(1,0)(1,1) depending on the resultant triggered momentary sensors in the exercise arm. This in turn produces a resultant output aligning to the expectant result in the defined truth table shown in (Fig 1(d)) and (Fig 2(a)).

Fig 2(e) details the activation of circuit 2 (colour 'yellow' active) (conductive thread) interconnections based on the bending of the elbow on the control arm of the soft exosuit and the inputs (0,1)(1,0)(1,1) into the circuit 2 coming from the triggered sensors in the exercise arm based on the exercise movements of the wearer.

### Dynamic FP Fabric Soft Exosuit Circuit Design Method

A layered, modular and interchangeable design method was adopted as a best approach towards the implementation of the dynamic FP soft Exosuit as seen in (Fig 4(a-0)). Here we detail the three layered design approach adopted which include 1) layer 1: Power and ground bus connections 2) layer 2: interchangeable embedded textile interconnections implemented with conductive thread and 3) layer 3: consists of modular and interchangeable fabric logic gate patches.

**Power and Ground Bus Connections**: When implementing a combinational logic circuit with varied logic gates, there is a necessity to include power (+VCC) and ground (GND) connections to support the operational functionality of the overall circuit and the individual logic gates. Due to the need for multiple power and ground connections a conductive thread embroidered bus network circuit topology was adopted within layer 1 of the soft exosuits back panel to house these connections. Having a separate layer specifically to host the power and ground connections allows for extensibility and flexibility to add/remove connections as required depending on the applied combinational circuitry required if modifications were necessary in the future. All +VCC and GND connections required from layer 3 (logic gate patches) could be connected using stainless steel snap fasteners that were capable of conducting the required power from one layer to the next based on the overall circuit requirements.

**Interchangeable Embedded Textile Interconnections Panel**: Layer 2 of the dynamic FP fabric consists of an interchangeable panel layer with conductive thread circuit interconnections. Circuit 1 clustered connections 2, 4 coming from the combinational circuit in the back panel of the soft exosuit are linked directly with the wrist sensor in the controlling right arm of the exosuit. Circuit 2 clustered connections 1, 3, 5 as seen in (Fig 1(b)) coming from the back panel of the soft exosuit are linked directly with the elbow triggered sensor in the controlling right arm of the exosuit. For each conductive thread interconnection point required to be connected between layer 2 or layer 3 (logic gate modular patches), we embedded snap fasteners at placement points. The control arm has 10 output and input snap fastener connections to the momentary sensors in the control right arm while the exercise arm left arm has 2 input snap fastener connections that were directly interconnected to the momentary exercise triggered sensors in the soft exosuit.

**Layer 3 Modular Logic Gate Patches:** Layer 3 involves the capability to exchange and replace the textile logic circuit gates in a dynamic programmable modular manner inter-weaved into layer 2 of the overall FP circuit. The modular textile logic gates (AND, OR and NOT) were fabricated using conductive fabric nylon, conductive thread and SMD resistor and transistor devices. Snap fasteners were utilised to interconnect the logic gate patches to power and ground bus connections in the soft exosuit. Through the creation of modular and interchangeable gate patches this provides an added dimension to the overall dynamic FP textile circuit as follows.

- For the current FP combinational logic circuit, logic gate patches can be removed and replaced. If a logic gate is deemed faulty and not operating as expected, replacement of the logic gate can easily be accomplished by actively removing the patch from the snap connectors and replacing it with a workable new logic gate patch.
- Operational functionality of the current FP combinational logic circuit in the fabric exosuit can also be modified depending on the type of logic gate patches utilised in conjunction with the layer 2 wired interconnection circuit. Hence, changing from an AND gate to an OR gate, the functionality and logic truth table of the circuit can be altered. This allows for a modifiable circuit adaptable to varying applications and use cases through the capability to alter the resultant outputs (logic truth table) applicable to the soft exosuit. This makes the overall design and implementation more dynamic, reusable and sustainable.

**Momentary Textile Sensors and Soft Exosuit Field Programmability**

Momentary based textile sensors were integrated at key placement points in the garment to act as actuators as well as dynamic FP circuit inter-changers, triggered by flexion and extension movements of the wearers elbow and wrist (controlling right arm of the garment). A total of six momentary textile sensors were created, 4 for the exercise arm and 2 for the control arm of the soft exosuit. Each textile momentary sensor was created using 2 outer layers of stretch Lycra fabric material with conductive fabric nylon tape strips and an internal middle layer consisting of non conductive Lycra (4 way stretch) fabric.

Movement through selected bending (wrist or elbow) of the soft exosuit control arm (right arm) in a flexion motion triggered the required circuit (1 or 2) to become operational in the exosuit. Depending on the activated circuit specific exercises could then be initiated on the exercise left arm of the soft exosuit. Such exercises ( as seen in Figure 2(b)) when triggered in the exercise arm, direct input values (aligning to the defined truth table as seen in (Fig 2(a))) to the operational FP combinational circuit in the exosuit. Based on the activation of the embedded momentary sensors interconnected to the dynamic FP circuit in the garment it is possible to define a set and personalised exercise program for the soft exosuit wearer. As per our defined upper limb post stroke rehabilitation we focused on the exercises defined in (Fig 1(d)) and (Fig 2(b)).

Integration and use of other textile sensor types ( eg pressure sensors, temperature sensors, electrodes ) in the dynamic FP soft exosuit could be another research extension beyond this initial experimentation phase extending the functionality and usability of the soft exosuit in a fabric environment. This is beyond the scope of this paper but will be investigated in future research.

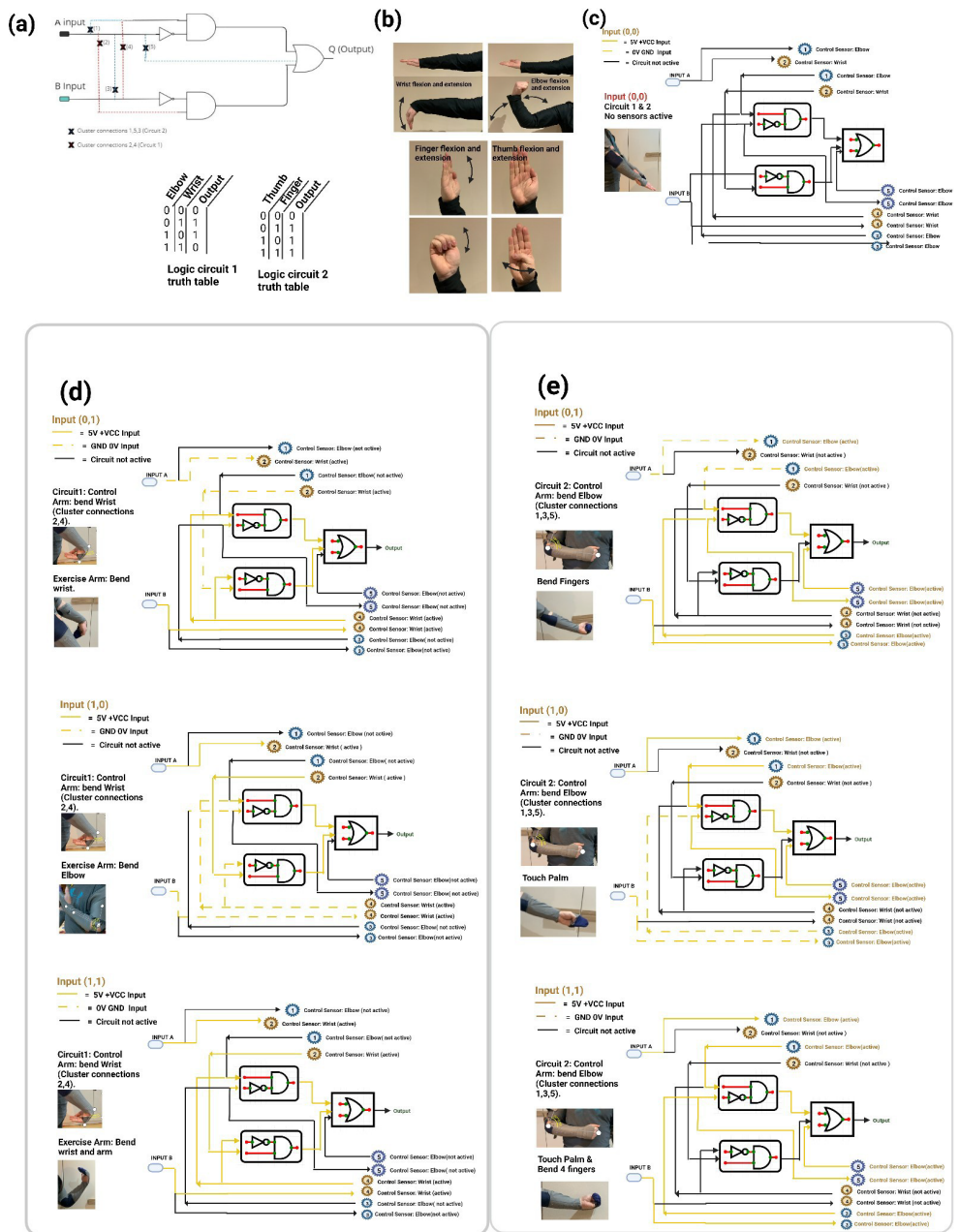

Fig. 2. a) Merged representation of both circuit 1 and circuit 2 with supporting logic truth tables highlighting the applied sensor exercise movements. b) range of exercise movements applicable to the embedded field programmable circuit. c)standard calibration (0,0) input wired connections representing zero input/zero output for both circuits. d)graphical representations of embedded circuit 1 active and non active wired conductive thread bus connections in the fabric soft exosuit. e)graphical representations of embedded circuit 2 active and non active wired conductive thread bus connections in the fabric exosuit.

### Healthcare Use Case and Workflow

In this paper we focus on a specific post stroke related healthcare use case linked to upper limb rehabilitation exer- cises and demonstrate the application of a dynamic FP fabric exosuit [23]. World Stroke Organisations (WSO) global stroke fact sheet 2022 states that over the timeframe of a year more than *12.2 million* new patients with strokes occur globally (one every 3 seconds) It is stated that one in four people over twenty-five will have a stroke in their lifetime, with over 62 percent of all stokes occurring in people under 70 (stroke is no longer considered a disease of the elderly) [24].

The *Brunnstroms* 7 stages of recovery after a stroke provides guidance for medics dealing directly with stroke patients highlighting the main recovery stages along with key mobility and exercises guidelines to help restore a patients motor function [25] [26] [27]. Post stroke rehabilitation exercises include upper limb strength training. Such exercises often focus on arm and hand exercises in order to retrain the brain and improve the patient's quality of life [28] [29]. Repetitive basic exercises of the wrist, arm and elbow 2 to 3 times a day, repeating each exercise 5-10 times or as advised following a stroke), help to gradually over time increase a patients mobility and flexibility. Upper limb and hand rehabilitation exercises have been selected for our use case based on physiotherapy stroke guidelines and arm exercises. Below is a list of the key simulated exercises we will utilise to demonstrate an application of the dynamic FP logic-driven fabric exosuit.

- Elbow Extension and Flexion: Bend and also straighten the simulatd patients elbow.
- Wrist Flexion and Extension: Bend wrist forward and backward.
- Finger Flexion and extension: The simulated patient bends their fingers to make a fist and then straightens then fully.
- Thumb Flexion and extension: Bending the thumb away and towards the base of the palm of the little finger.

Most post-stroke patients require an individually tailored exercise program depending on the level of rehabilitation required, hence having a wearable smart garment with in- terchangeable application intelligence is vital to support the personalised approach required [25] [30].

## DESIGN AND IMPLEMENTATION

This section provides an overview of the design and imple- mentation of the dynamic FP logic-driven fabric soft exosuit, highlighting the design structure of the garment and embedded operational functionality.

### *Layered Dynamic Field Programmable Circuit Implementation*

Introducing modularity and interchangeability of the dy- namic FP circuit from a design perspective brings an element of sustainability, re-usability and flexibility to the embedded intelligence in the fabric exosuit, all the while enabling easy modification of the garment, its embedded circuitry and ap- plication use cases. Implementation of the design focused on a shrug soft exosuit pattern. The shrug soft exosuit style was chosen as it was deemed a better option of a garment design for a user (post stroke with limited motion and movement capability) to put on and take off with ease, rather than the option of a compression type garment that had to be placed over the head and arms and required alot of effort to place on the wearer. This shrug design was modified to incorporate soft exosuit element designs (flexibility/velcro straps at arm/joint movement areas) to produce a more usable and adaptable soft exosuit prototype.

To enable both movement and structure in the soft exosuit fabric design, specific types of fabrics were selected for use. A *Scuba Lycra 2 way stretch (grey colour)* fabric was selected to fabricate the main panels of the shrug jacket garment. In contrast to allow for enhanced movement in the garment elbow momentary textile sensors placement points in the arms of the garments, a *4 way stretch lycra (navy colour)* material was used, this type of fabric provided additional flexibility required ensuring close body-fit and ease of placement of the textiles sensors ensuring enhanced performance and usability as illustrated in (Fig 4(a-g)).

From a pattern design perspective using the above stated material fabric types, a combination of a shrug/soft exosuit design pattern combined with an *arm gauntlet design pattern* were utilised in order to create a bespoke pattern design for the dynamic FP fabric exosuit. For the creation of the power (+VCC), ground (GND) and logic gate interconnection bus lines in the textile fabric garment environment (back panel of the garment seen in (Fig 4(h)), a conductive thread *Madeira HC12* was utilised in the circuit creation process. Surface mount devices (SMD) *0805 resistor 1k Ohm and 0805 resistor 10k Ohm* resistors as well as *2N2222 SMD NPN Transistors* were utilised in the creation of interchangeable logic gate patches (*AND,OR and NOT gates*) building on previous textile logic gates research [31] [32] [33].To enable fabric patch interchangeability and modularity in the soft exosuit *snap*

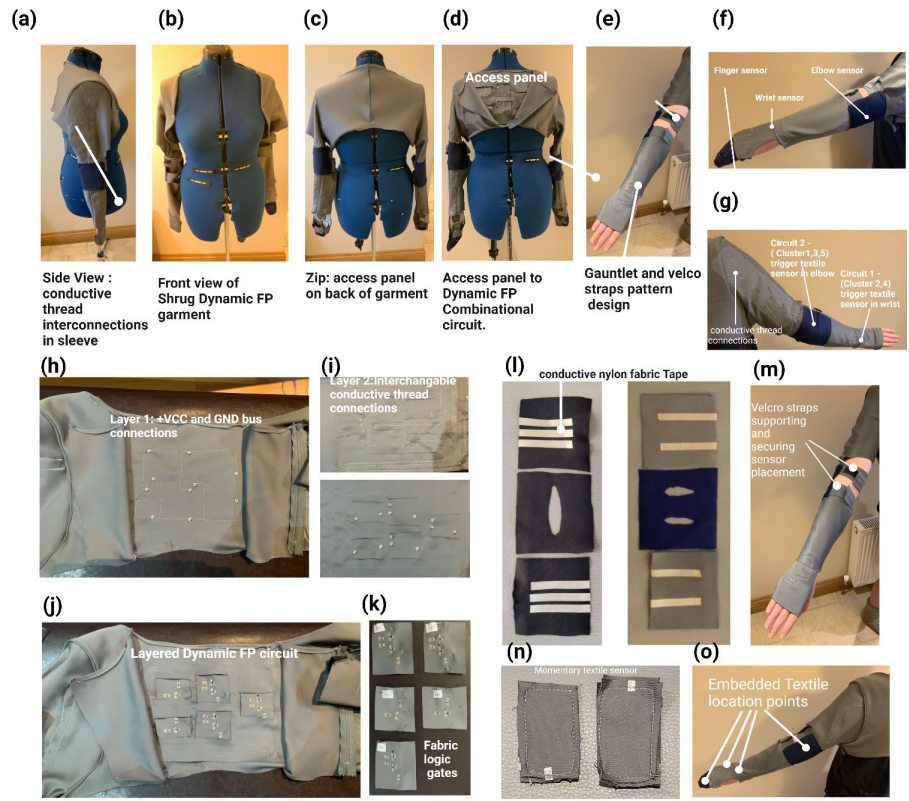

Fig. 4. a-g) 360 degrees visuals showcasing the fabric soft exosuit h) Layer 1 Power (+VCC) and GND bus connections embroidered into the garment base layer i) Layer 2 Circuit interconnections interchangeable layered pat, j) Combined 3 layers interconnected in the shrug smart garment back panel k) Modular and interchangeable Logic gate patches(AND,OR,NOT). l) Visual of momentary textile sensors on lycra based fabric with fabric nylon tape m) Velcro strap design above and below the sensor integration at the elbow points allowing for manual manipulation and placement of the sensors int motion placement points depending on the individuals arm shape and size n-o ) Momentary textile sensors and location placement points on the exercise arm of the shrug smart jacket.

*fasteners 9mm* were used in the garment allowing modular fabric intelligence patches to be inserted, interchanged and removed as required. The following details some of the main key design elements that were implemented.

- The back panel of the shrug jacket houses dynamic FP layered fabric based circuit as discussed in section 2. Due to the layered nature of the dynamic FP circuit, an access panel and zip were included to allow for easy access to the FP fabric circuit allowing replacement of the modular logic gate patches and the combinational conductive thread interconnected layered patch if required as illustrated in (Fig 4(c-d)).
- For the soft exosuit arm design, due to the placement of the textile momentary sensors at key areas of the body (wrist, elbow, fingers) flexibility and mobility were an essential consideration as shown in (Fig 4(e-g)). At each elbow joint on both the control and exercise arm, the textile momentary sensors were integrated into the 4 way stretch navy lycra fabric of the soft exosuit. To allow for enhanced movement of the elbow joint and precise placement of the momentary sensor at the required point on the elbow, it was decided to apply velcro straps above and below the elbow textile sensor region, enabling flexibility of placement as well as the option to secure the textile sensors at a given location using the extendable velcro straps (depending on the arm size requirements of the individual) illustrated in (Fig 4(e)(m)).

- A gauntlet style lower arm fabric pattern design was adopted providing a close fit to the arm enabling en- hanced mobility performance for the textile sensors as seen in (Fig 4(e)).
  - Circuit connections to the arm textile sensors were em- bedded into the arms of the garment using a zigzag sewing machine stitch as illustrated in (Fig 4(a)) and (Fig 4(g)). Conductive thread *Madeira HC12* was used to create these interconnections in the fabric environment. Scuba and lycra type fabrics are difficult to work with when sewing on a machine and using conductive thread due to the movability in the fabric. To provide stability to the design when sewing the zigzag stitching, an inter- facing layer was used on the inside layer of the fabric that made up the arm of the shrug jacket. Snap fasteners were used to connect the arm sensor interconnections to the dynamic FP circuit on the back panel of the garment.
  - *The control arm - right arm of smart garment* as seen in (Fig 4(g)) contains 2 key control momentary textile sensors, one triggering circuit 1 located in the wrist of the gauntlet and one triggering circuit 2 located in the elbow (navy 4 way stretch Lycra material).
  - *The exercise arm - left arm of smart garment* as seen in (Fig 4(f)) contains 4 momentary textile sensors located in the elbow, wrist, fingers and palm of hand.

To note in future iterations of this prototype, further investigation into the designs the arms of the soft exosuit to make them more modular and interchangeable will be investigated, allowing the option to switch exercises from left to right hand side of the garment. This is out of scope for this initial experimental prototype.

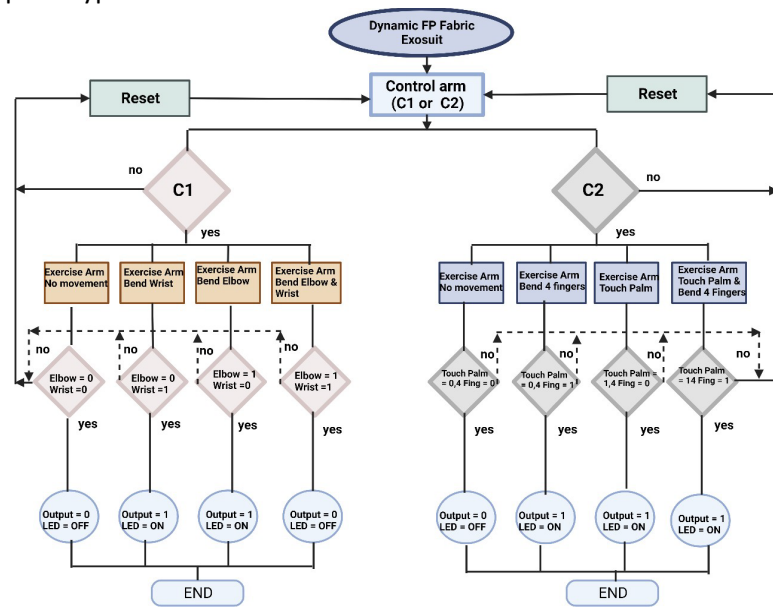

Fig. 3. Dynamic Field Programmable Fabric Exosuit Flowchart.

Taking these identified rehabilitation exercises a corresponding dynamic FP soft exosuit process flowchart model was designed and illustrated in (Fig 3). The model of the soft exosuit, provides the end users (technical and non-technical) with an understanding of the garments functionality and operational processes. The model conveys the control points placed internally within the garment as well as the flow of the embedded data, triggered based on the movement of the control arm and exercise arm of the soft exosuit. The sequential steps of the flowchart details the process in order to document and visualise the concept prior to progressing to the design and implementation phase.

## RESULTS

Taking the created soft exosuit prototype, it was necessary to verify the operational functionality of the dynamic FP soft exosuit. This involved the following test phases

- Verify the operational functionality of the combinational logic circuits to ensure the logic gate functionality is operating as expected and aligns to the defined truth tables in (Fig 1(d)).
- Verify the dynamic and interchangeable field programma- bility capability of the fabric exosuit to switch from *circuit 1 to circuit 2* as required triggered by the relevant momentary sensors in the control arm of the garment.
- Assessment of triggered iterative movement exercises using the garment, while also taking into account what is considered normal range of motion (ROM) and align- ing to the currently used post stroke Fugl-Meyer (FM) assessment scoring scale.
- Demonstration of the potential of the fabric exosuit to perform everyday task focused exercise driven scenarios.

This section presents the results obtained from the validation experiments focusing on the operational functionality of the dynamic FP logic driven soft exosuit observing the actual results obtained with the expected outputs.

### *Logic - Driven FP Combinational Circuit Gate Validation*

Core to the dynamic FP soft exosuit is the operation of its embedded logic gates. In the defined overall dynamic FP combinational circuit illustrated in (Fig 1(a)(b)) we have a mixture of gates including *AND, OR and NOT* logic gates. It was necessary to check the operational logic of each gate based on the input values (*A,B*) to the circuit. A set input voltage of *5 Volts* +VCC was used for an input representing a logic '*1*' and GND represented an input of logic '*0*'. For output results obtained, a voltage reading above *2 Volts* (set threshold) was deemed to represent an *output logic '1'* and a voltage reading below *2 Volts* represented *output logic'0'*. Each truth table input combination was initiated three times and the voltage measurement at each logic gate in the circuit was measured and recorded. This was completed for both circuit 1 and circuit 2. Fig 5(a)(b) visually details the output voltage readings obtained.

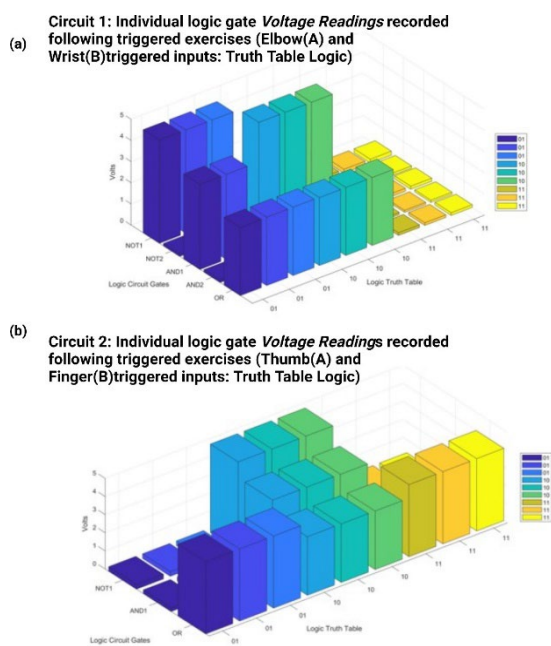

Fig. 5. a) Combinational Logic Gate Circuit 1 measurements at each individual logic gate level. b) Combinational Logic Gate Circuit 2 voltage measurements at each individual logic gate level.

From the results obtained we can state the following voltage measurement observational readings for each of the gates as defined in (Fig 2(a)). Voltage measurements obtained aligned to the expected output values as per the logic truth tables seen in (Fig 2(a)) and Table 1. When tested the readings remained stable with no extreme voltage fluctuations detected. On two occasions irregular readings were detected, upon further inspection a frayed conductive thread connection from layer 2 to the *AND1* logic gate SMD transistor was detected and was causing the transistor to malfunction. Once corrected and the short eliminated, this issue was resolved. The second issue encountered was a result of a broken solder connection from the conductive fabric nylon tape to the *SMD 10K Ohm* resistor of the *AND2* logic gate in the combinational circuit. Once re-soldered, this resolved the connectivitiy issue and the circuit operated as expected.

### TABLE I

#### LOGIC GATE CIRCUIT VOLTAGE READINGS

| Circuit: Gate | Voltage readings |
|---|---|
| **Circuit 1** *Input* (0, 1) | NOT1(4.88V),NOT2)118.6mV), AND1(4.02V),AND2(119.4mV), OR(3.23V) |
| **Circuit 1** *Input* (1, 0) | NOT1(237.9mV),NOT2(4.91V), AND1(196.5mV),AND2(62.9mV), OR(3.21V) |
| **Circuit 1** *Input* (1, 1) | NOT1(183.4mV),NOT2(106.6mV), AND1(141.6mV),AND2(96.5mV), OR(123.6mV) |
| **Circuit 2** *Input* (0, 1) | NOT1(255mV),AND1(213.9mV), OR(3.98V) |
| **Circuit 2** *Input* (1, 0) | NOT1(4.89V),AND1(4.02V), OR(3.23V) |
| **Circuit 2** *Input* (1, 1) | NOT1(290.6mv),AND1(247.8mV), OR(3.99V) |

*Soft Exosuit Exercise Angle Movement Analysis*

In order to assess the dexterity of the soft exosuit, we investigated the garments *Elbow* and *Wrist* goniometry (range of motion). We considered normal range of motion in the overall movement angular measurements tests [34].

The following details the normal ranges of movement for the

1) Elbow: Flexion(150 degrees), Prunation (80 degrees) Supination (80 degrees).
2) Wrist: Flexion(60 degrees), Extension(60 degrees), Ab- duction(20 degrees)
3) Finger joints (Interphalangeal proximal (PIP) middle knuckles of fingers): Flexion (120 degrees) Exten- sion(120 degrees).
4) Thumb (interphalangeal (middle knuckle) joint): Flexion (80 degrees), Extension (90 degrees).

Following three repetitive exercise movements as per each of the assigned exercises highlighted in (Fig 6(a)), using a clinometer we were able to record the angle measurements (degrees) of each exercise. When an exercise movement was initiated, once the LED (connected to the FP embedded cir- cuitry in the soft exosuit) appeared in an '*ON*'state, movement of the arm ceased and an image and angle measurement in degrees was taken. This was repeated *3 times* for each exercise as seen in (Fig 6(a)). This provided us with a threshold angle level of dexterity that aligned to the momentary sensor triggers in the exercise arm of the soft exosuit. Based on the measurements recorded we can state that the average wrist flexion recorded a *52 degrees* measurement, the elbow flexion averaged a *107 degree* measurement, the thumb flexion measured on average *69 degrees* and finally the finger flexion registered an average of *88 degrees*. To note this is the angle at which the selected embedded combinational circuit configuration was active, as well as the selected momentary touch senors located in the exercise arm. The prototype does have the capability to extend beyond these recorded readings up to the the normal range of motion highlighted at the start of this section.

### Dynamic Field Programmable Soft Exosuit Analysis

As per most patient focused rehabilitation exercise pro- grams, it is necessary to complete multiple repetitive iter- ations of assigned exercises in order to see a progression and improvement in the overall range of motion over time. In order to validate the performance of the dynamic FP soft exosuit, repetitive exercise movements (*10 repetitive movement exercises*) aligning to the defined truth table illustrated in (Fig 2(d)), were completed over a set time-frame in order to assess the usability and performance of the prototype. Assessment of the FP circuit switching operational functionality during the exercise program was important to ensure dynamic reconfig- uration and the soft exosuit application operated as expected. To support this validation experiment, a *LilyPad Accelerometer (ADXL335)* was also attached to the exercise arm gauntlet front side of the hand. This enabled an additional range of motion based datasets to be obtained during this specific test, linked to the soft exosuits prototype defined exercise range of motion (elbow, wrist, finger and thumb flexion).

Accelerometry data correlates with movement tests demon- strating the upper limb range of motion and can contribute insights into day to day activities of relevance that help in the overall assessment of an individual's recovery post stroke. Within this experiment, we gathered raw accelerometry data that enabled a tri-dimensional representation to support the distinction and data spatial pattern for the various exercises completed. Using the range of motions, angle degree mea- surements from the soft exosuit (using clinometer) and the output LED alert notification in the garment, along with the accelerometry data obtained, we were capable of scoring the results of the soft exosuit data, aligning it to a subscale of the Fugl-Meyer (FM) assessment scale (Fig 7). The FM assess- ment scale measures and evaluates key recovery domains in post-stroke patients. It covers 5 key domains (Motor function, Sensory function, balance, joint range of motion, joint pain). This FM assessment scale is widely used in both research and clinical environments currently. Aligning to our use case, we therefore focused on the FM sub-scale domain that assess joint range of motion and joint pain, linked specifically to the upper extremity. Items to be scored aligning to a 2-point ordinal scale *1=performs partially, 2=performs fully*. Taking this FM scoring for range of motion we mapped it to the angle measurements of the soft exosuit that demonstrate a full motion and partial motion also aligned to the LED alert notification of the soft exosuit LED highlighted by the state of the LED. LED '*ON*' demonstrator full motion, LED *OFF*' state demonstrating partial motion as seen in (Fig 8).

*Experiment setup protocol*: The momentary touch sensor embedded placement points in the soft exosuit (elbows, wrist, fingers and thumb) can be cross checked to ensure alignment to the wearers movement joints. If required, slight manual manoeuvring of the sensors embedded in the 4 way stretch fabric is possible enabling realignment. Once placement points of the soft exosuit are aligned, the velcro straps (located at the elbow points on the soft exosuit) are secured. This adds a layer of stability to the design and structure when exercises were being performed. Validation of a zero input, zero output static state calibration is completed. This involves placing the individuals two arms straight down by their side

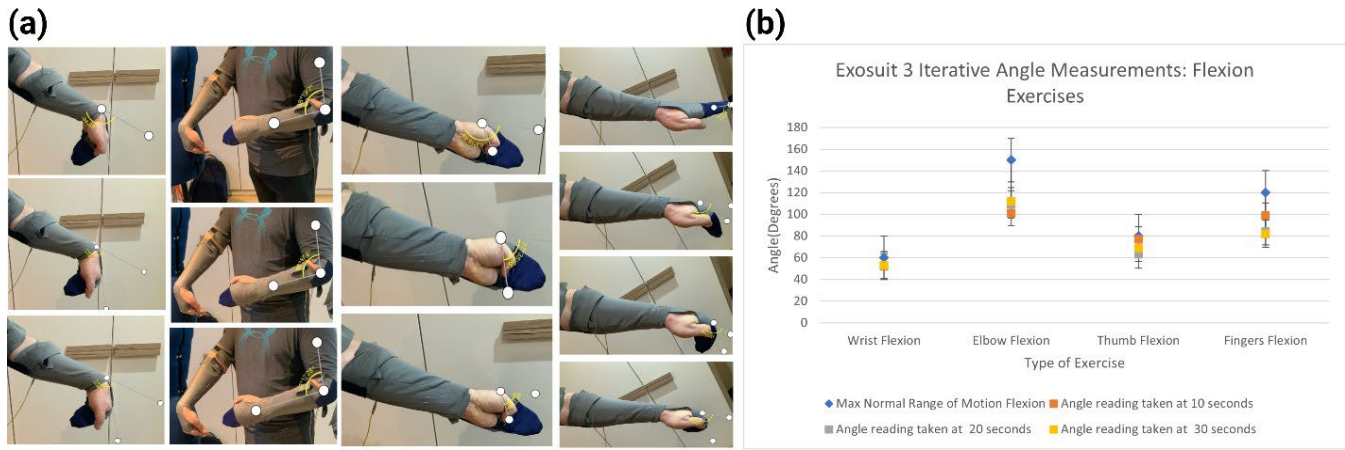

Fig. 6. a) Iteractive image snapshots of repetitive exercise movement b) Angle movements recorded demonstrating the angle measurements obtained are within the normal main range of motion

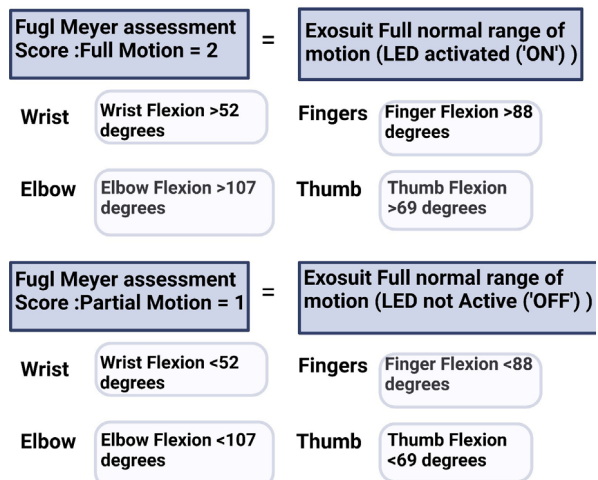

Fig. 7. Mapping Fugl-Meyer (FM) assessment to exercises performed using the dynamic soft exosuit (angle measurements) through alignment with LED activation and non activation states

and ensuring the LED light linked to the output of the Dynamic FP embedded fabric circuit was in an '*OFF*'state.

*Experiment execution*: 10 repetitive upper limb exercises were completed within an overall set time-frame of between *100 to 150* seconds. The following main exercises that trig- gered specific inputs/outputs depending on the Dynamic FP logic-driven embedded circuit activated were completed.

- Control Arm (Bend Wrist), Exercise Arm (Wrist Flex- ion (10 repetitive movements)
- Control Arm (Bend Wrist), Exercise Arm (Elbow Flex- ion (10 repetitive movements) )
- Control Arm (Bend Elbow), Exercise Arm (Finger Flexion (10 repetitive movements))
- Control Arm (Bend Elbow), Exercise Arm (Thumb Flexion (10 repetitive movements))

*Experiment result:* During the exercises, datasets and mea- surements were recorded. For each individual exercise per- formed two graphical representations of the movements were completed as seen in (Fig 8). A voltage output measurement coming from the combinational circuit activated was recorded to demonstrate the momentary sensors triggering and the dynamic switching of circuits as seen in (Fig 8). The results were plotted to visually demonstrate the voltages recorded against time, once each exercise was initiated and repeated 10 times. From the results we observe for *circuit 1 wrist flexion* an output 'ON' state has a voltage fluctuation between *2.8 Volts and 2.91 Volts*. For *circuit 1 elbow flexion* input we observed a voltage fluctuation of between *2.85 Volts and 3.03 Volts* for an output 'ON' state from the dynamic FP logic- driven exosuit. In comparison for *circuit 2* when active the finger flexion demonstrated a voltage output range of between *1.97 Volts to 2.54 Volts* while the thumb flexion output voltage ranged from *2.61 Volts to 2.67 Volts.* as seen in (Fig 8).

Due to our set threshold of above 2 Volts determining an *ON* and exercise complete state, we observed in the Finger flexion exercise group, an output dipping to *1.97 Volts* as seen in (Fig 8), which was slightly below the defined set threshold, hence due to the LED being in state '*OFF*' a partial movement FM assessment was mapped to this output result. Upon inspection of the momentary sensor embedded in the soft exosuit, it was noted that the output connection from the sensor that acted as an input trigger to the FP circuit in the soft exosuit, had become frayed and this was causing an inconsistency and fluctuation in its operations state. This conductive thread connection was replaced and its continuity crosschecked using a digital multi-meter to ensure no shorts remained. From the results obtained in (Fig 8) through the observational LED alerts during the exercises along with the tridimensional 3D scatter plot patterns demonstrating the movement of the exer- cises and the Voltage output results showcasing the triggered circuitry operating demonstrating the dynamic FP. Such 3D scatter plots convey a spatial representation of the acceleration movement initiated and recorded when the exercises were completed. Data obtained with positive values demonstrate an increase in velocity, nagative values demonstrate a decrease in velocity. Based on analysis of the x,y,z datasets (negative and positive values), we can identify the clustering of dataset values that shows the flexion and extension movement of the exosuit when completing its exercises as highlighted in (Fig 8).

Fig. 8. a (i-iii))Wrist Flexion movements completed 10 times, graphs demonstrating accelerometer 3D scatter patterns per each exercise, voltage output levels from the dynamic FP circuit and fugl meyer mapped assessment scores (motion and pain) based on the soft exosuit movements for each exercise.

A mapping of the results obtained in could be aligned to the FM assessment scale and a scoring recorded (Fig 8). Further future assessment of this mapping is required in a follow on future human trials where a larger group of participants and the results recorded can provide further and more in-depth results showcase the alignment of the dynamic soft exosuit to the FM assessment scoring method.

## SOFT EXOSUIT TASK DRIVEN APPLICATION SCENARIOS

For stroke rehabilitation patients it is vital to understand the principle of neuroplasticity and the brains capability to regain lost function [35]. Research has demonstrated that regular rehabilitation upper limb exercises linked to daily activities and tasks can be more impactful and effective than none task related upper limb and hand exercises [36]–[38]. Retraining the brain and learning a new skill requires hours of repetitive exercise. Post stroke remote exercise programs need to incorporate task-driven rehabilitation training. The Dynamic FP soft exosuit provides the option to incorporate such daily task-driven exercises into the wearers scheduled home training program.

The following provides an insight into an example of such activities.

1) Task: Lift Bag as seen in (Fig 9(a)), this task driven exercise involved placement of the control arm be- tween on average *78 to 98 degrees* to enable *circuit 2* dynamically. This in turn enables the operation of the '*Thumb*' and '*Finger*' flexion momentary sensors in the exercise arm of the soft exosuit. Once the bag was lifted this the Finger flexion sensor triggered an input of *(0,1)* and an output of *'1'* activating the LED to demonstrate successful completion of the exercise. Depending on the method of lifting the bag, a wrist extension of an angle averaging *24 degrees*, while also bending the fingers could also be obtained, providing additional mobility practice beyond the set finger flexion exercise.

2) Task: Lift Bottle as seen in (Fig 9(b)) With the control arm elbow momentary sensor activated by placing the control arm elbow of the soft exosuit in approximately *78 to 98 degree* angle, this in turn activates the task specific exercise involving lifting a green bottle up from a table and placing it back down onto the table was completed. Through enabling the elbow sensor of the control arm this in turn triggers the use of the thumb momentary sensor on the palm of the garment of the exercise hand (left arm of the garment). When lifting the bottle up and down from the table in an repetitive execise, the bottle directly comes in contact with the thumb palm momentary sensor to validate the exercise (LED in '*ON*' state).

3) Task: Hold Rolled towel as seen in (Fig 9(c)) Using a small rolled towel, this is the perfect object to help practice and repeat wrist flexion exercises. This specific task aligns to using the wrist momentary sensors embedded at the wrist placement point on the exercise arm of the soft exosuit. A flexion angle of ranging between *49 to 58 degrees* was measured demonstrating the flexion angle that can be achieved through this task driven exercise. For the control arm an average angle of *55 degrees* triggered the dynamic FP change of the embedded intelligence to operate at circuit 1 level in the garment.

4) Task: Elbow flexion while holding can of food. To utilize the embedded momentary sensor in the elbow of the exercise arm, a task driven exercise involving bending the elbow while holding a can of food in the exercise arm as illustrated in (Fig 9(d)). On av- erage an angle of *88 degrees* triggered the successful completion of the elbow task based exercise. Such an exercise using elbow based motion can be applied across many task driven scenarios.

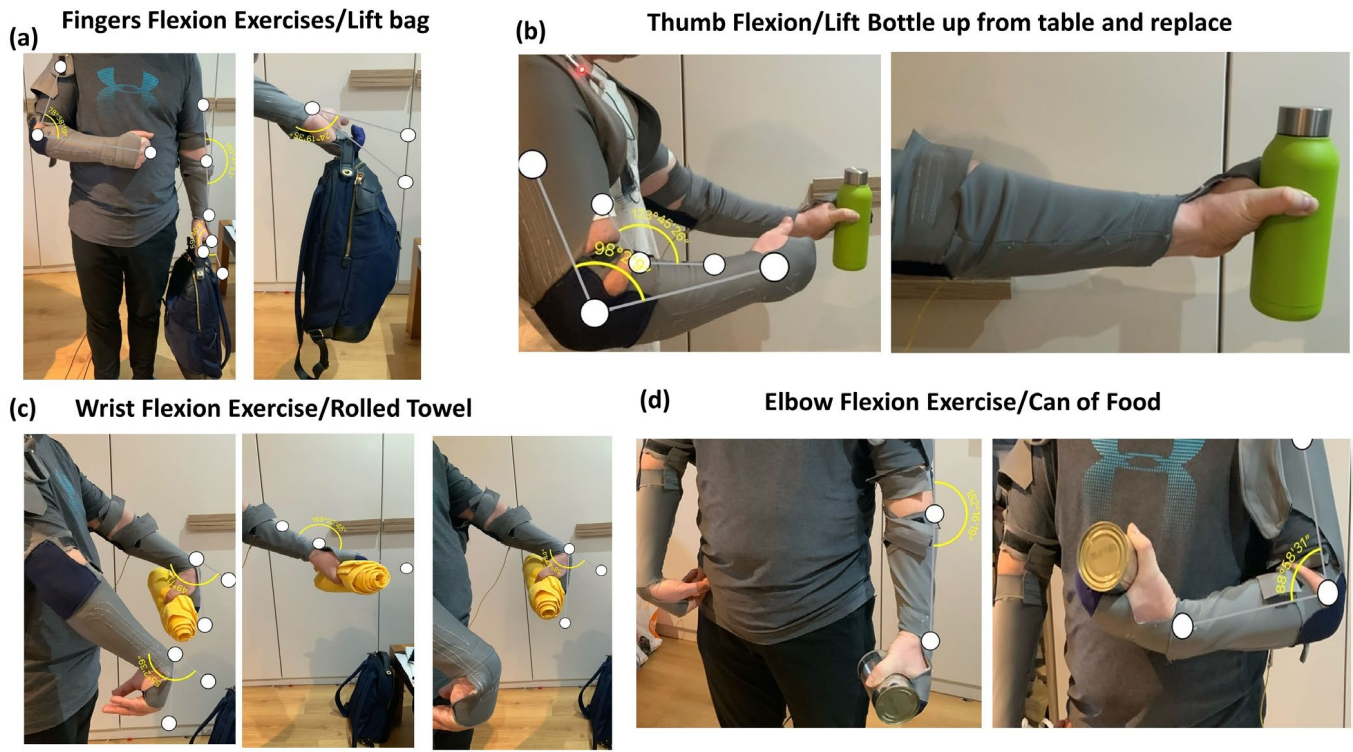

Fig. 9. a) Finger flexion motion exercises for everyday task of lifting a bag b) Thumb flexion motion exercises to lift a bottle up from the table and replace it back on the table c) Wrist Flexion movement angles utilising a small towel to support the activity and link to a daily task d) Flexion of the elbow daily task using a can of food

## CONCLUSION

Fabric computing, intelligent fabric, advanced functional fibres are terms in the etextiles and smart material domain that are starting to gain traction and new innovative etextiles ideas are emerging around what *an intelligent fabric space* and *fabric computing wearable* of the future might look like [39] [40] [1]. Contributing towards this overarching and forward-looking vision, this paper focuses on the design and implementation of a dynamic FP logic-driven soft exosuit prototype capable of embedded circuit reconfiguration functionality.

To achieve this we successfully designed and implemented an initial working prototype taking inspiration from field programmability where the internal hardware can be modified without disassembling the device and combinational logic circuitry where input signals applied have an immediate effect on the circuits output providing real-time resultant outputs for suitable applications. We implemented an innovative soft exosuit prototype that had the capability to dynamically alter its embedded circuitry configuration and operational state triggered through motion based movements of the wearers wrist and arm (control arm, right arm of the soft exosuit).

To support the validation of the soft exosuit we focused on a healthcare rehabilitation use case linked to post-stroke rehabilitation upper limb exercise needs to demonstrate the soft exosuits potential. Due to the layered, modular and interchangeable logic gate patch and interconnection patch approach, this further extends the usability, adaptability and flexibility of the Dynamic FP soft exosuit to be adaptable and usable across multiple applications and scenarios. Key findings of the study include

- Creation of a dynamic field programmable logic-driven soft exosuit enabling embedded reconfiguration and operational functionality of the soft exosuit, triggered and controlled via the wearers right arm elbow and wrist movements.

- Design and development of a modular and layered embedded and interchangeable FP combinational circuit based approach in a soft exosuit fabric environment.
- Design and implementation of logic gate patches utilising surface mount devices (SMD), conductive thread and conductive fabric.
- Demonstration and validation of a healthcare focused post stroke rehabilitation use case linked to upper limb rehabilitation exercises

This initial experimental prototype demonstrated in this paper drives innovation around the application of dynamic field programmability in a textile environment. Such an innovation also seeks to enhance etextile sustainability through the adopting of modular interchangeable functionality hence increasing the potential application and reuse of such an application which in turn contributes towards prolonging the life-cycle of the soft exosuit and reducing textile waste. Future research areas of focus include

- Scalability of the dynamic field programmable embedded logic circuit to support a larger number of inputs and outputs.
- Further investigation around the progression towards a more complete textile embedded circuit solution, replacing SMD hard devices with textile equivalents where possible (eg textile resistors).
- Creation of dedicated modular embroidery patterns, that will include set machine embroidery stitching and guidelines for the creation of specific modular combi- national logic circuitry enabling field programmability and that can be embedded easily through the embroi- dery and conductive thread into a textile environment adopting the patch based approach that is interchange- able and replaceable.
- Further investigation around the generation of a next iteration of the prototype that will incorporate varying other textile sensor types (eg pressure sensors, tem- perature sensors, electrode sensors) to support other biometric readings.
- Further development of the prototype design and struc- ture to incorporate the possibility to alternate the exercise and control functionalities of the soft exosuit between its two arms hence extending the capability of the soft exosuit and enhancing its flexibility from a rehabilitation application point of view.
- Future research planning to validate the soft exo- suit prototype from an end user perspective, through planned human trials to further investigate future itera- tions of the prototype to assess usability, performance and durability.

## ACKNOWLEDGMENT

B. C Henshall is funded in part by FutureNeuro from Science Foundation Ireland (SFI) under Grant Number 16/RC/3948 and co-funded under the European Regional De- velopment Fund and by FutureNeuro industry partners. Select Figures are created with BioRender.com FC, SB, and DH devised the project focus, the main conceptual ideas, and the proof outline. FC worked on the technical development elements and validation of the designs. FC, SB, WS and DH proposed the validation plan and contributions. FC wrote the manuscript with input from authors SB, DH and WS.

## CONFLICT OF INTEREST

The authors declare that the research was conducted in the absence of any commercial or financial relationships that could be construed as a potential conflict of interest.

## FUNDING

DH is funded in part by FutureNeuro from the Science Foundation Ireland (SFI) under grant number *16/RC/3948* and cofunded under the European Regional Development Fund and by FutureNeuro industry partners

## ETHICS STATEMENT

Written informed consent for publication has been obtained from the subject, for any non- identifiable human images to be published in the Journal.

## REFERENCES

[1] D. C. C¸ elikel, "Smart e-textile materials," *Advanced Functional Mate- rials*, pp. 1–16, 2020.

[2] M. Ahsan, S. H. Teay, A. S. M. Sayem, and A. Albarbar, "Smart clothing framework for health monitoring applications," *Signals*, vol. 3, no. 1, pp. 113–145, 2022.

[3] S. C. Sethuraman, P. Kompally, S. P. Mohanty, and U. Choppali, "My- wear: a novel smart garment for automatic continuous vital monitoring," *IEEE Transactions on Consumer Electronics*, vol. 67, no. 3, pp. 214–222, 2021.

[4] M. Janusz, M. Roudjane, D. Mantovani, Y. Messaddeq, and B. Gosselin, "Detecting respiratory rate using flexible multimaterial fiber electrodes designed for a wearable garment," *IEEE Sensors Journal*, vol. 22, no. 13, pp. 13 552–13 561, 2022.

[5] Y. Zheng, N. Tang, R. Omar, Z. Hu, T. Duong, J. Wang, W. Wu, and H. Haick, "Smart materials enabled with artificial intelligence for healthcare wearables," *Advanced Functional Materials*, vol. 31, no. 51, p. 2105482, 2021.

[6] S. u. Zaman, X. Tao, C. Cochrane, and V. Koncar, "Smart e-textile systems: A review for healthcare applications," *Electronics*, vol. 11, no. 1, 2022.

[7] K. Yang, B. Isaia, L. J. Brown, and S. Beeby, "E-textiles for healthy ageing," *Sensors*, vol. 19, no. 20, 2019.

[8] C. Massaroni, J. Di Tocco, M. Bravi, A. Carnevale, D. Lo Presti, R. Sabbadini, S. Miccinilli, S. Sterzi, D. Formica, and E. Schena, "Respiratory monitoring during physical activities with a multi-sensor smart garment and related algorithms," *IEEE Sensors Journal*, vol. 20, no. 4, pp. 2173–2180, 2020.

[9] M. I. M. Esfahani and M. A. Nussbaum, "A "smart" undershirt for tracking upper body motions: Task classification and angle estimation," *IEEE Sensors Journal*, vol. 18, pp. 7650–7658, 2018.

[10] J. Rose, A. El Gamal, and A. Sangiovanni-Vincentelli, "Architecture of field-programmable gate arrays," *Proceedings of the IEEE*, vol. 81, no. 7, pp. 1013–1029, 1993.

[11] V. B. K. L. Aruna, C. Ekambaram, and M. Padmaja, "Field pro- grammable gate array implementation of an adaptive filtering based noise reduction and enhanced compression technique for healthcare applications," *Transactions on Emerging Telecommunications Technolo- gies*, vol. 34, no. 1, p. e4654, 2023.

[12] "Field-programmable biological circuits and configurable (bio)logic blocks for distributed biological computing," *Computers in Biology and Medicine*, vol. 128, p. 104109, 2021.

[13] M. R. Azghadi, C. Lammie, J. K. Eshraghian, M. Payvand, E. Donati, B. Linares-Barranco, and G. Indiveri, "Hardware implementation of deep network accelerators towards healthcare and biomedical applications," *IEEE Transactions on Biomedical Circuits and Systems*, vol. 14, no. 6, pp. 1138–1159, 2020.

[14] F. Cleary, W. Srisa-an, B. Gil, J. Kesavan, T. Engel, D. C. Henshall, and S. Balasubramaniam, "Wearable fabric brain enabling on-garment edge- based sensor data processing," *IEEE Sensors Journal*, vol. 22, no. 21, pp. 20 839–20 854, 2022.

[15] R. Sanchez-Iborra and A. F. Skarmeta, "Tinyml-enabled frugal smart objects: Challenges and opportunities," *IEEE Circuits and Systems Magazine*, vol. 20, no. 3, pp. 4–18, 2020.

[16] P. Manickam, S. A. Mariappan, S. M. Murugesan, S. Hansda, A. Kaushik, R. Shinde, and S. P. Thipperudraswamy, "Artificial intel- ligence (ai) and internet of medical things (iomt) assisted biomedical systems for intelligent healthcare," *Biosensors*, vol. 12, no. 8, 2022.

[17] K. Dong, X. Peng, and Z. L. Wang, "Fiber/fabric-based piezoelectric and triboelectric nanogenerators for flexible/stretchable and wearable electronics and artificial intelligence," *Advanced Materials*, vol. 32, no. 5, p. 1902549, 2020.

[18] M. F. Waheed and A. M. Khalid, "Impact of emerging technologies for sustainable fashion, textile and design," pp. 684–689, 2019.

[19] G. Loke, J. Alain, W. Yan, T. Khudiyev, G. Noel, R. Yuan, A. Missakian, and Y. Fink, "Computing fabrics," *Matter*, vol. 2, no. 4, pp. 786–788, 2020.

[20] https://www.globenewswire.com/en/search/organization/Absolute, "Electronic textiles market size shares by 2028 revenue, cost analysis, gross margins, future investment segmentation by types, applications key players, market dynamics," November 2022.

[21] C. H. Roth Jr, L. L. Kinney, and E. B. John, *Fundamentals of logic design*. Cengage Learning, 2020.

[22] C. Chen, Z. Wang, W. Li, H. Chen, Z. Mei, W. Yuan, L. Tao, Y. Zhao, G. Huang, Y. Mei, Z. Cao, R. Wang, and W. Chen, "Novel flexible material-based unobtrusive and wearable body sensor networks for vital sign monitoring," *IEEE Sensors Journal*, vol. 19, no. 19, pp. 8502–8513, 2019.

[23] R. van der Vliet, R. W. Selles, E.-R. Andrinopoulou, R. Nijland, G. M. Ribbers, M. A. Frens, C. Meskers, and G. Kwakkel, "Predicting upper limb motor impairment recovery after stroke: a mixture model," *Annals of neurology*, vol. 87, no. 3, pp. 383–393, 2020.

[24] V. L. Feigin, M. Brainin, B. Norrving, S. Martins, R. L. Sacco, W. Hacke, M. Fisher, J. Pandian, and P. Lindsay, "World stroke organization (wso): global stroke fact sheet 2022," *International Journal of Stroke*, vol. 17, no. 1, pp. 18–29, 2022.

[25] S. Brunnstrom, "Motor Testing Procedures in Hemiplegia: Based on Sequential Recovery Stages," *Physical Therapy*, vol. 46, no. 4, pp. 357–375, 04 1966.

[26] S. Naghdi, N. N. Ansari, K. Mansouri, and S. Hasson, "A neurophysi- ological and clinical study of brunnstrom recovery stages in the upper limb following stroke," *Brain Injury*, vol. 24, no. 11, pp. 1372–1378, 2010, pMID: 20715900.

[27] C.-Y. Huang, G.-H. Lin, Y.-J. Huang, C.-Y. Song, Y.-C. Lee, M.-J. How, Y.-M. Chen, I.-P. Hsueh, M.-H. Chen, and C.-L. Hsieh, "Improving the utility of the brunnstrom recovery stages in patients with stroke: Validation and quantification," *Medicine*, vol. 95, p. e4508, 08 2016.

[28] A. P. Salazar, C. Pinto, J. V. Ruschel Mossi, B. Figueiro, J. L. Lukrafka, and A. S. Pagnussat, "Effectiveness of static stretching positioning on post-stroke upper-limb spasticity and mobility: Systematic review with meta-analysis," *Annals of physical and rehabilitation medicine*, vol. 62, no. 4, p. 274—282, July 2019.

[29] J. M. Rondina, C.-h. Park, and N. S. Ward, "Brain regions important for recovery after severe post-stroke upper limb paresis," *Journal of Neurology, Neurosurgery & Psychiatry*, vol. 88, no. 9, pp. 737–743, 2017.

[30] C. J. Winstein, J. Stein, R. Arena, B. Bates, L. R. Cherney, S. C. Cramer, F. Deruyter, J. J. Eng, B. Fisher, R. L. Harvey *et al.*, "Guidelines for adult stroke rehabilitation and recovery: a guideline for healthcare profession- als from the american heart association/american stroke association," *Stroke*, vol. 47, no. 6, pp. e98–e169, 2016.

[31] F. Cleary, D. C. Henshall, and S. Balasubramaniam, "On-body edge computing through e-textile programmable logic array," *Frontiers in Communications and Networks*, vol. 2, 2021.

[32] A. Rajappan, B. Jumet, R. A. Shveda, C. J. Decker, Z. Liu, T. F. Yap, V. Sanchez, and D. J. Preston, "Logic-enabled textiles," *Proceedings of the National Academy of Sciences*, vol. 119, no. 35, p. e2202118119, 2022.

[33] I. Posch and E. Kurbak, "Crafted logic towards hand-crafting a com- puter," in *Proceedings of the 2016 CHI Conference Extended Abstracts on Human Factors in Computing Systems*, 2016, pp. 3881–3884.

[34] J. M. SOUCIE, C. WANG, A. FORSYTH, S. FUNK, M. DENNY, K. E. ROACH, D. BOONE, and T. H. T. C. NETWORK, "Range of motion measurements: reference values and a database for comparison studies," *Haemophilia*, vol. 17, no. 3, pp. 500–507, 2011.

[35] M. Dimyan, "Neuroplasticity in the context of motor rehabilitation after stroke," *Nature reviews. Neurology*, vol. 7, pp. 76–85, 02 2011.

[36] A. M. Alsubiheen, W. Choi, W. Yu, and H. Lee, "The effect of task- oriented activities training on upper-limb function, daily activities, and quality of life in chronic stroke patients: A randomized controlled trial," *International Journal of Environmental Research and Public*

*Health*, vol. 19, no. 21, 2022.

[37] C. Yoo and J. Park, "Impact of task-oriented training on hand function and activities of daily living after stroke," *Journal of physical therapy science*, vol. 27, no. 8, pp. 2529–2531, 2015.

[38] W. Choi, "The effect of task-oriented training on upper-limb function, visual perception, and activities of daily living in acute stroke patients: A pilot study," *International Journal of Environmental Research and Public Health*, vol. 19, no. 6, p. 3186, 2022.

[39] M. Chen, J. Liu, P. Li, H. Gharavi, Y. Hao, J. Ouyang, J. Hu, L. Hu, C. Hou, I. Humar, L. Wei, G.-Z. Yang, and G. Tao, "Fabric computing: Concepts, opportunities, and challenges," *The Innovation*, vol. 3, no. 6, p. 100340, 2022.

[40] M. Chen, R. Wang, Y. Zhou, Z. He, X. Liu, M. He, J. Wang, C. Huang, H. Zhou, P. Hong *et al.*, "Digital medical education empowered by intelligent fabric space," *National Science Open*, vol. 1, no. 1, p. 20220011, 2022.